%% file: Lattice2017_354_RIGGIO.tex
\documentclass[epj]{webofc}
\usepackage[utf8]{inputenc}
\usepackage[varg]{txfonts}   
\usepackage{booktabs}
\usepackage{xcolor}
\definecolor{darkred}{rgb}{0.4,0.0,0.0}
\definecolor{darkgreen}{rgb}{0.0,0.4,0.0}
\definecolor{darkblue}{rgb}{0.0,0.0,0.4}
\usepackage[bookmarks,linktocpage,colorlinks,
linkcolor = darkred,
urlcolor  = darkblue,
citecolor = darkgreen]{hyperref}

\usepackage{array} 

\usepackage{amsmath,amssymb,mathrsfs,textcomp,amscd,bbm} 

\usepackage{graphicx} 
\DeclareGraphicsExtensions{.pdf,.png,.jpg,.mps,.ps,}
\graphicspath{{pics/}}

\usepackage{braket}

\usepackage{subfigure}
\usepackage[rightcaption]{sidecap}
\wocname{EPJ Web of Conferences}
\woctitle{Lattice2017}
\input{definitions.tex}

\begin{document}

%
\selectlanguage{english}
\title{%
	Tensor form factor for the $D \to \pi(K)$ transitions with Twisted Mass fermions.\thanks{presented at Lattice 2017 - 35th International Symposium on Lattice Field Theory, 18th-24th June 2017, Granada.}
}
\author{%
\firstname{Vittorio} \lastname{Lubicz}\inst{1,2} \and
\firstname{Lorenzo}  \lastname{Riggio}\inst{1} \and
\firstname{Giorgio} \lastname{Salerno}\inst{1,2} \and
\firstname{Silvano} \lastname{Simula}\inst{1} \and
\firstname{Cecilia} \lastname{Tarantino}\inst{1,2}
}
\institute{%
INFN, Sezione di Roma Tre, Rome, Italy
\and
Dipartimento di Matematica e Fisica, Universit\'a di Roma Tre, Rome, Italy
}
\abstract{%
  We present a preliminary lattice calculation of the $D \to \pi$ and $D \to K$ tensor form factors $f_T(q^2)$ as a function of the squared 4-momentum transfer $q^2$. ETMC recently computed the vector and scalar form factors $f_+(q^2)$ and $f_0(q^2)$ describing $D \to \pi(K) \ell \nu$ semileptonic decays analyzing the vector current and the scalar density. The study of the weak tensor current, which is directly related to the tensor form factor, completes the set of hadronic matrix element regulating the transition between these two pseudoscalar mesons within and beyond the Standard Model where a non-zero tensor coupling is possible. 
Our analysis is based on the gauge configurations produced by the European Twisted Mass Collaboration with $N_f = 2 + 1 + 1$ flavors of dynamical quarks. We simulated at three different values of the lattice spacing and with pion masses as small as 210 MeV and with the valence heavy quark in the mass range from $\simeq 0.7\, m_c$ to $\simeq 1.2\, m_c$. The matrix element of the tensor current are determined for a plethora of kinematical conditions in which parent and child mesons are either moving or at rest. As for the vector and scalar form factors, Lorentz symmetry breaking due to hypercubic effects is clearly observed in the data. We will present preliminary results on the removal of such hypercubic lattice effects.
}
\maketitle
\section{Introduction and simulation details}\label{intro}

Precise measurements of hadron weak decays can constrain the Standard Model (SM) and place bounds on New Physics (NP) models.
The transitions between pseudoscalar (PS) mesons can be parametrized, in all extensions of the SM, in terms of three form factors, namely $f_+$, $f_0$ and $f_T$. New physics from heavy particles such as those appearing in models with supersymmetry, a fourth generation, or composite Higgs sectors alter Wilson coefficients in the effective Hamiltonian that describes physics below the electroweak scale. Whatever these unknown particles may be, the hadronic physics remains the same.

Recently in Ref.~\cite{Lubicz:2017syv}, we presented the first $N_f = 2+1+1$ lattice QCD (LQCD) calculation of the vector and scalar form factors $f_+^{D \to \pi(K)}(q^2)$ and $f_0^{D \to \pi(K)}(q^2)$ governing the semileptonic $D \to \pi(K) \ell \nu$ decays, using the gauge configurations generated by the European Twisted Mass Collaboration (ETMC) with $N_f = 2 + 1 + 1$ dynamical quarks, which include in the sea, besides two light mass-degenerate quarks, also the strange and charm quarks with masses close to their physical values \cite{Baron:2010bv,Baron:2011sf}. in this contribution we complete the set of operators relevant for the $D \to \pi(K)$ transitions analyzing the weak tensor current $\bar{c} \sigma_{\mu\nu} q$ to extract  $f_T^{D \to \pi(K)}(q^2)$. 
The latter one may enter as a BSM contribution both in semileptonic $D \to \pi(K) \ell \nu$ decays and in rare decays $c \to u \ell^+ \ell^-$, which are loop-soppressed in the SM as they proceed through flavor-changing neutral currents.

Similarly to Ref.~\cite{Lubicz:2017syv}, we have evaluated tensor form factor in the whole accessible range of values of $q^2$ in the experiments, i.e.~from $q^2 = 0$ up to $q^2_{\rm max} = (M_D - M_{\pi(K)})^2$.
In our calculations quark momenta are injected on the lattice using non-periodic boundary conditions \cite{Bedaque:2004kc,deDivitiis:2004kq} and the matrix elements of the tensor current are determined for many kinematical conditions, in which parent and child mesons are either moving or at rest.
The data exhibit a remarkable breaking of Lorentz symmetry due to hypercubic effects for both $D\to \pi$ and $D\to K$ form factors. The presence of these effects has already been observed in Ref.~\cite{Lubicz:2017syv} for the vector and scalar form factors, and in that paper we presented a method to subtract the hypercubic artefacts and recover the Lorentz-invariant form factors in the continuum limit.
Apart from Ref.~\cite{Lubicz:2017syv}, hypercubic effects have never been observed in the context of the $D \to \pi(K)$ transitions. Previous lattice calculations, however, used only a limited number of kinematical conditions (typically the $D$-meson at rest). We argue that this may obscure the presence of hypercubic effects in the lattice data. 
These effects appear to be affected by the difference between the parent and the child meson masses. This is clearly a very important issue, which warrants further investigations. If this is the case, these effects will play an important role in the determination of the form factors governing semileptonic $B$-meson decays into lighter mesons and is therefore crucial to have them under control.

In \cite{Lubicz:2017syv} the subtraction of the observed hypercubic effects was achieved by considering a more general decomposition of the matrix element, which contains, beside the usual Lorentz-covariant part, additional hypercubic form factors proportional to $a^2$. The form of this decomposition is related to the Lorentz structure of the current in the matrix element. It is therefore interesting, to further validate the method, to apply it also in the case of the tensor current. In this contribution we present the subtraction of hypercubic artefacts and the determination of the tensor $D \to \pi(K)$ form factors.

The gauge ensembles and the simulations used in this work are the same adopted in Ref.~\cite{Lubicz:2017syv} so we refer the interested reader to Sec.~2 of \cite{Lubicz:2017syv} for a more detailed discussion. Here we only want to stress that since we work with the Wilson Twisted Mass Action at maximal twist \cite{Frezzotti:2000nk,Frezzotti:2003xj,Frezzotti:2003ni}, an automatic ${\cal{O}}(a)$-improvement \cite{Frezzotti:2003ni,Frezzotti:2004wz} is guaranteed for our lattice setup.

The QCD simulations have been carried out at three different values of the inverse bare lattice coupling $\beta$, to allow for a controlled extrapolation to the continuum limit, and at different lattice volumes.
For each gauge ensemble we have used a number of gauge configurations corresponding to a separation of 20 trajectories to avoid autocorrelations.  
We have simulated quark masses in the range from $\simeq 3\, m_{ud}$ to $\simeq 12\, m_{ud}$ in the light sector, from $\simeq 0.7\, m_s$ to $\simeq 1.2\, m_s$ in the strange sector, and from $\simeq 0.7\, m_c$ to $\simeq 1.2\, m_c$ in the charm sector, where $m_{ud}$, $m_s$ and $m_c$ are the physical values of the average up/down, strange and charm quark masses respectively, as determined in Ref.~\cite{Carrasco:2014cwa}.
The lattice spacings are found to be $a = \{ 0.0885\,(36), 0.0815\,(30), 0.0619\,(18)\}\, \fm$ at $\beta = \{1.90, 1.95, 2.10\}$, respectively, the lattice volume goes from $\simeq 2$ to $\simeq 3$ fm and the pion masses, extrapolated to the continuum and infinite volume limits, range from $\simeq 210$ to $ \simeq 450\,\MeV $.

\section{Lattice calculation of the tensor matrix elements and hypercubic effects}\label{sec-1}

The matrix element of the tensor current $T_{\mu \nu}$ between an initial $D$-meson state and a $\pi$($K$)-meson final state can be written, as required by the Lorentz symmetry, in terms of a single form factor $f_T(q^2)$:
\be
    \left\langle {T^{\mu \nu}} \right\rangle \equiv \left\langle {P(\,{p_P})} \right|{T^{\mu \nu}}\left| {D(\,{p_D})} \right\rangle  = \frac{2}{{{M_D} + {M_P}}}\left[ {p^\mu _P p^\nu _D - p^\nu _P p^\mu _D} \right]{f_T}\left( {{q^2}} \right) ~ ,
    \label{eq:matrix_element_decomposition}
\ee
where $P = \pi (K)$ can be either the pion or the kaon and the 4-momentum transfer $q$ is given by $q \equiv p_D - p_P$ and the factor ${M_D} + {M_P}$ is conventionally inserted to make the form factor dimensionless. 

In order to inject momenta on the lattice we impose non-periodic boundary conditions (BC's) \cite{Bedaque:2004kc,deDivitiis:2004kq,Guadagnoli:2005be} using the setup described in Refs.~\cite{Lubicz:2017syv,Carrasco:2016kpy} for the $D \to \pi(K)$ semileptonic decays and the $K_{\ell 3}$ decays.
The $\pi$, $K$ and $D$ meson 3-momenta are then given by $p = \frac{2 \pi}{L} \left( \theta , ~ \theta  ,~ \theta \right)$, where the parameter $\theta$, democratically distributed along the three spatial directions, can assume for each gauge ensemble the values collected in Table~3 of Ref.~\cite{Lubicz:2017syv}.
These values have been chosen in order to obtain momenta with values ranging from $\approx 150\MeV$ to $\approx 650\MeV$ for all the various lattice spacings and volumes. 

In our lattice calculation we make use of the local version of the tensor current $\bar{c} \sigma _{\mu \nu} q$ with $q = d, s$. Since we employ maximally twisted fermions, the tensor current renormalizes multiplicatively \cite{Frezzotti:2003ni}, i.e.~$\widehat{T}_{\mu \nu} = {{\cal{Z}}_T} \bar{c} \sigma _{\mu \nu} q$. The tensor renormalization constant (RC) ${\cal{Z}}_T$ has been computed in the RI$^\prime$-MOM scheme by using dedicated ensembles of gauge configurations produced with $N_f = 4$ degenerate flavors of sea quarks~\cite{Carrasco:2014cwa}. Two different methods, labelled as M1 and M2, were employed and they are expected to lead to the same final results once the continuum limit for the physical quantity of interest is performed. The numerical values of ${{\cal{Z}}_T}$ converted in the $\overline{\rm MS}$ scheme are: $Z_T^{\overline{\rm MS}}(2\,\mathrm{GeV})(M_1) = \{ 0.711\,(5), 0.724\,(4), 0.774\,(4)\}$ and $Z_T^{\overline{\rm MS}}(2\,\mathrm{GeV})(M_2) = \{ 0.700\,(3), 0.711\,(2), 0.767\,(2)\}$ at $\beta = \{ 1.90, 1.95, 2.10 \}$.

Since the tensor current can be written in terms of one form factor only, even a single component of $\widehat{T}_{\mu \nu}$ is in principle sufficient to extract $f_T$. We have evaluated $\braket{\widehat{T}_{0i}} \equiv {{\cal{Z}}_T}\left\langle {P(\,{p_P})} \right|  \bar{c} \sigma _{0i} q \left| {D(\,{p_D})} \right\rangle$ ($i=1, 2, 3$), with the first and second indexes corresponding to temporal and spatial components, respectively.
Furthermore, since we are using democratically distributed momenta in the three spatial directions, the matrix elements of the tensor current $\braket{\widehat{T}_{01}}$, $\braket{\widehat{T}_{02}}$ and $\braket{\widehat{T}_{03}}$ are equal to each other. 
Therefore, in order to improve the statistics, we average them to get
 \be
       \label{eq:Tsp}
       \braket{\widehat{T}_{i}} \equiv \frac{1}{3} \left[ \braket{\widehat{T}_{01}} + \braket{\widehat{T}_{02}} + 
                                                           \braket{\widehat{T}_{03}} \right]~.
 \ee
 
The matrix elements $\braket{\widehat{T}_{0i}}$ can be extracted from the large (Euclidean) time distance behavior of a convenient combination of 2- and 3-point correlation functions in lattice QCD, which are defined as 
\bea
   \label{eq:C2}
           \quad
   C_2^{D(P)}\left(t^\prime,\,\vec{p}_{D(P)}\right)\!\!\!\! & = & \!\!\!\!\frac{1}{L^3} \sum_{\vec{x},\vec{z}} \braket{0 \lvert P_5^{D(P)}(x)\,P_5^{D(P)\dagger}(z) 
          \rvert 0}\,e^{-i\vec{p}_{D(P)}\cdot(\vec{x}-\vec{z})}\,\delta_{t^\prime,\,(t_x-t_z)}~,\\
   \label{eq:C3}
   C^{DP}_{\widehat{T}_{0i}}\left(  t,\, t^\prime,\, \vec{p}_D,\, \vec{p}_P \right)\!\!\!\! & = & \!\!\!\!\frac{1}{L^6}\sum_{\vec{x},\vec{y},\vec{z}} 
           \braket{0\lvert P_5^{P}(x)\widehat{T}_{0i}(y)\,P_5^{D\dagger}(z)\rvert0}\,e^{-i\vec{p}_{D}\cdot(\vec{y}-\vec{z}) + i\vec{p}_{P}\cdot(\vec{y}-\vec{x})}\,
            \delta_{t,\,(t_y-t_z)}\,\delta_{t^\prime,\,(t_x-t_z)}~,\,\,\quad
\eea
where $t^\prime$ is the time distance between the source and the sink, $t$ is the time distance between the insertion of the tensor current and the source, while $P_5^D = i\,\bar{c} \gamma_5 u$ and $P_5^{\pi(K)}=i\, \bar{d} (\bar{s}) \gamma_5 u$ are the interpolating fields of the $D$ and $\pi(K)$ mesons. 

As is well known, at large time distances 2- and 3-point correlation functions behave as
 \bea
        \label{eq:C2_larget}
        \qquad \quad  \quad
        C_2^{D(P)}\left(t^\prime,\,\vec{p}_{D(P)}\right) \!\!\!\!& ~ _{\overrightarrow{t^\prime \gg a}} ~ &\!\!\!\! \frac{|Z_{D(P)}|^2}{2E_{D(P)}} 
            \left[ e^{-E_{D(P)} t^\prime} + e^{-E_{D(P)} (T - t^\prime)} \right] , \\
        \label{eq:C3_larget}        
        C^{DP}_{\widehat{T}_{0i}}\left(  t,\, t^\prime,\, \vec{p}_D,\, \vec{p}_P \right) \!\!\!\!& ~ _{\overrightarrow{t\gg a\,, \, (t^\prime-t)\gg a}} ~ &\!\!\!\!  
            \frac{Z_P Z_D^*}{4E_P E_D}\, \braket{P(p_P)|\widehat{T}_{0i}|D(p_D)}\, e^{-E_D t}\, e^{-E_P (t^\prime - t)}~,
 \eea
where $Z_D$ and $Z_P$ are the matrix elements $\braket{0\lvert\,P_5^D(0)\,\rvert\,D(\vec{p}_D) }$ and $\braket{0\lvert\,P_5^P(0)\,\rvert\,P(\vec{p}_{P})}$, which depend on the meson momenta $\vec{p}_D$ and $\vec{p}_P$ because of the use of smeared interpolating fields (see Ref.~\cite{Lubicz:2017syv} for details), while $E_{D(P)}$ is the energy of the $D$($P$) meson. The matrix elements $Z_D$ and $Z_P$ can be extracted directly by fitting the corresponding 2-point correlation functions. 

The matrix elements $\langle \widehat{T}_{i} \rangle$ (see Eq.~(\ref{eq:Tsp})) can be extracted from the time dependence of the ratios $R$ of 2- and 3-point correlation functions, defined as in Eqs.~(\ref{eq:C2}-\ref{eq:C3}), namely
 \be
    \label{eq:Rmu} 
    R\left( t, t', \vec{p}_D, \vec{p}_P \right) = 4 E_D  E_P \frac{ C_{T_{i}}^{DP}
        \left( t, t', \vec {p}_D, \vec{p}_P \right) ~ C_{T_{i}}^{PD}\left( t, t', \vec{p}_D, \vec{p}_P \right) }
        { \tilde{C}_2^D\left( t', \vec{p}_D \right) ~ \tilde{C}_2^P\left( t', \vec{p}_P \right)} ~ ,
 \ee
where the correlation function $\widetilde{C}_2^{D(P)}(t)$ is given by
 \be
      \widetilde{C}_2^{D(P)} \left(t,\,\vec{p}_{D(P)}\right) \equiv \frac{1}{2}\left[ C_2^{D(P)}\left(t,\,\vec{p}_{D(P)}\right) + 
          \sqrt{C_2^{D(P)}\left(t,\,\vec{p}_{D(P)}\right)^2 - C_2^{D(P)}\left(\frac{T}{2},\,\vec{p}_{D(P)}\right)^2} \right] ~ ,
      \label{eq:C2_tilde}
 \ee
which at large time distances behave as
 \be
      \widetilde{C}_2^{D(P)} \left(t,\,\vec{p}_{D(P)}\right) ~ _{\overrightarrow{t \gg a}} ~ Z_{D(P)} ~
          e^{-E_{D(P)} t} /  (2 E_{D(P)}) ~ ,
     \label{eq:C2_tilde_larget}
 \ee
i.e.~without the backward signal. At large time distances one has
\be
\label{eq:R_plateau}
R(t, t', \vec{p}_D, \vec{p}_P) _{\overrightarrow{t \gg a, ~ (t^\prime-t) \gg a}} ~ \lvert\braket{P(p_P)| \widehat{T}_{i} |D(p_D)}\rvert^2 = \lvert\braket{\widehat{T}_{i}}\rvert^2 ~ .
\ee
Finally, we consider the two kinematics with opposite spatial momenta and perform the average
 \be
       \label{eq:improvement_Ti}
       \braket{\widehat{T}_{i}}_{\rm imp}  \equiv \frac{1}{2} \left[ \braket{P(E_P, \vec{p}_P) | 
           \widehat{T}_{i} | D(E_D, \vec{p}_D)} - \braket{P(E_P, -\vec{p}_P) | \widehat{T}_{i} | 
           D(E_D, -\vec{p}_D)} \right] ~ , 
\ee
which guarantees the ${\cal{O}}(a)$ improvement on the matrix elements for moving mesons \cite{Frezzotti:2003ni}.

\begin{figure}[htb!] 
\centering
\includegraphics[width=11cm,clip]{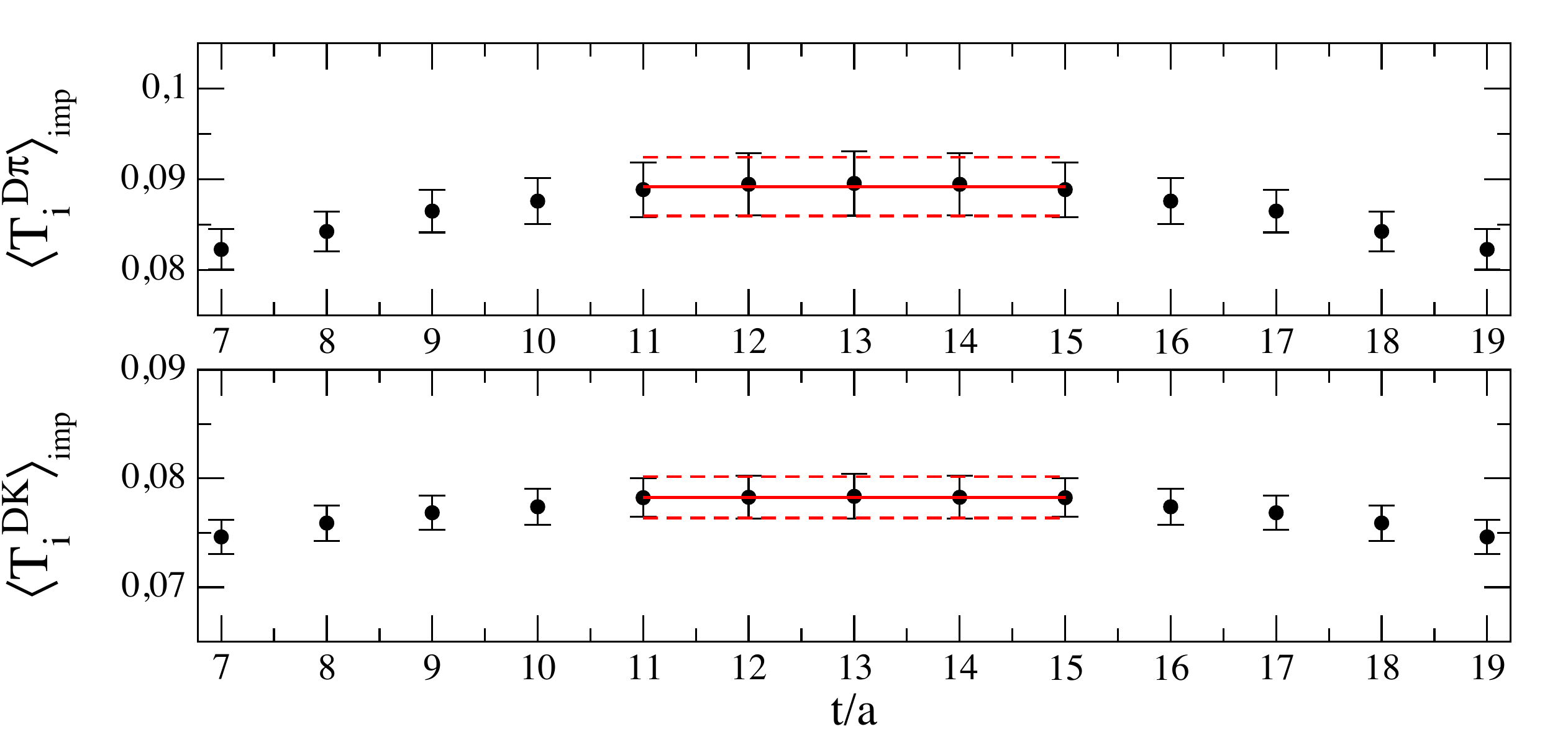}
\vspace{-0.4cm}
\caption{\it Matrix elements $\braket{\widehat{T}^{D \pi}_{i}}_{\rm imp}$ and  $\braket{\widehat{T}^{D K}_{i}}_{\rm imp}$ extracted from the ratio (\ref{eq:Rmu}) for the ensemble D20.48 with $\beta = 2.10$, $L / a = 48$, $\vec{p}_D = - \vec{p}_\pi$ and $|\vec{p}_D| \simeq 150\MeV$. The meson masses are $M_\pi \simeq 254 \MeV$, $M_K \simeq 516 \MeV$ and $M_D \simeq 1640\MeV $. The horizontal red lines correspond to the plateau regions used in the fit.}
\label{fig:plateau}
\end{figure}

The quality of the plateau for the matrix elements $\braket{\widehat{T}^{D \pi}_{i}}_{\rm imp}$ and  $\braket{\widehat{T}^{D K}_{i}}_{\rm imp}$ is illustrated in Fig.~\ref{fig:plateau}.
The time intervals adopted for fitting Eq.~(\ref{eq:R_plateau}) are $[t^\prime / 2 - 2, ~ t^\prime / 2 + 2]$ with the values of $t^\prime$ given in Table II of Ref.~\cite{Lubicz:2017syv}. They are compatible with the dominance of the $\pi$, $K$ and $D$ mesons ground-state observed for the two-point correlation functions.

Thus, from the 2- and 3-point lattice correlators we are able to extract the ${\cal{O}}(a)$-improved matrix elements $\braket{\widehat{T}_{i}}_{\rm imp}$. 
The standard procedure for determining the tensor form factor $f_T(q^2)$ is to assume the following Lorentz-covariant decomposition
 \be
     \label{eq:T_final}
     \braket{\widehat{T}_{i}}_{\rm imp} = \frac{2}{M_D + M_P}  
         \left[ E_P p^i_D - E_D p^i_P \right] f_T(q^2) + {\cal{O}}(a^2) ~ . 
  \ee

\begin{figure}[htb!] 
\centering
\makebox[\textwidth][c]{
\includegraphics[width=6.75cm,clip]{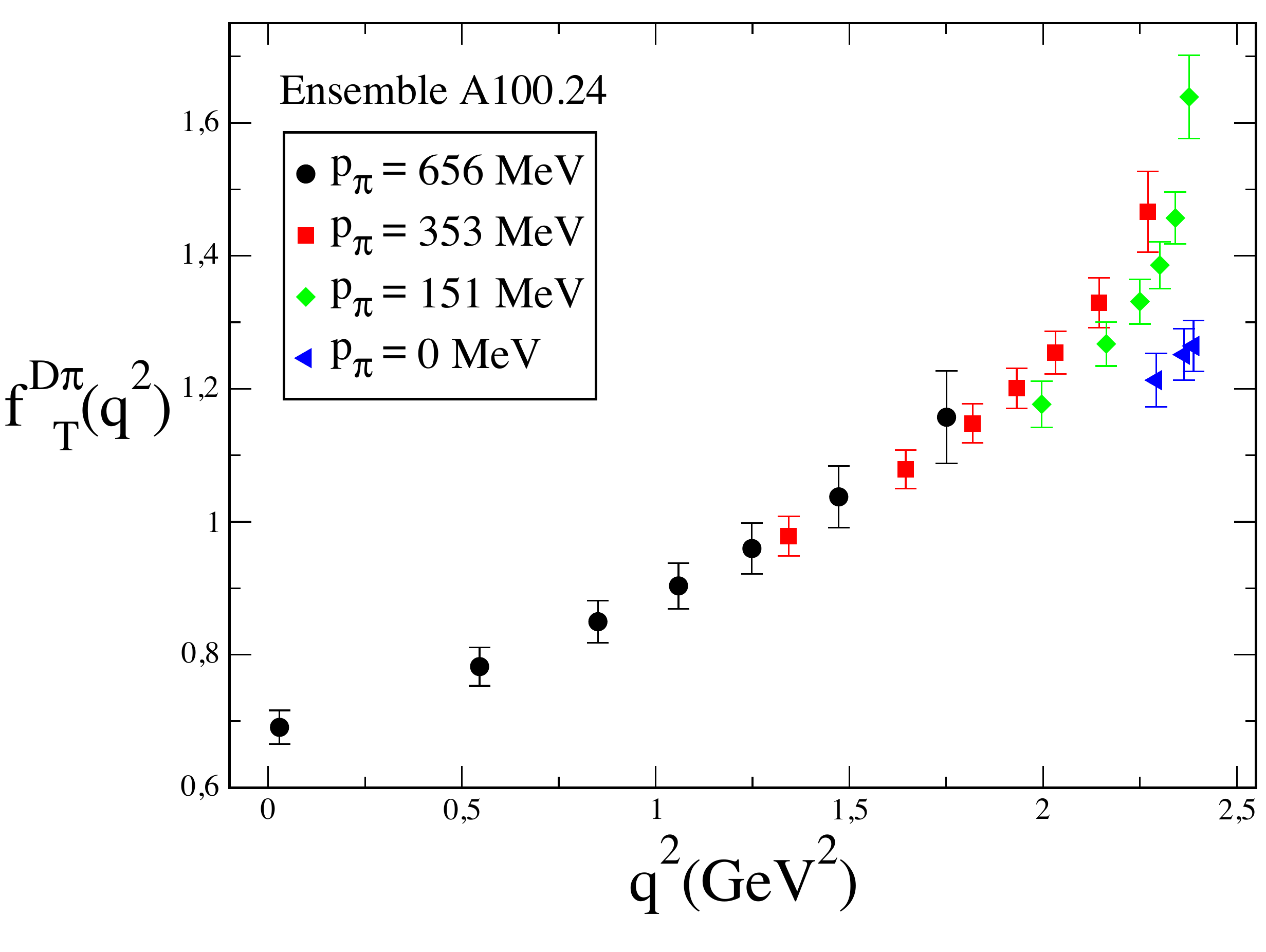}
\includegraphics[width=6.75cm,clip]{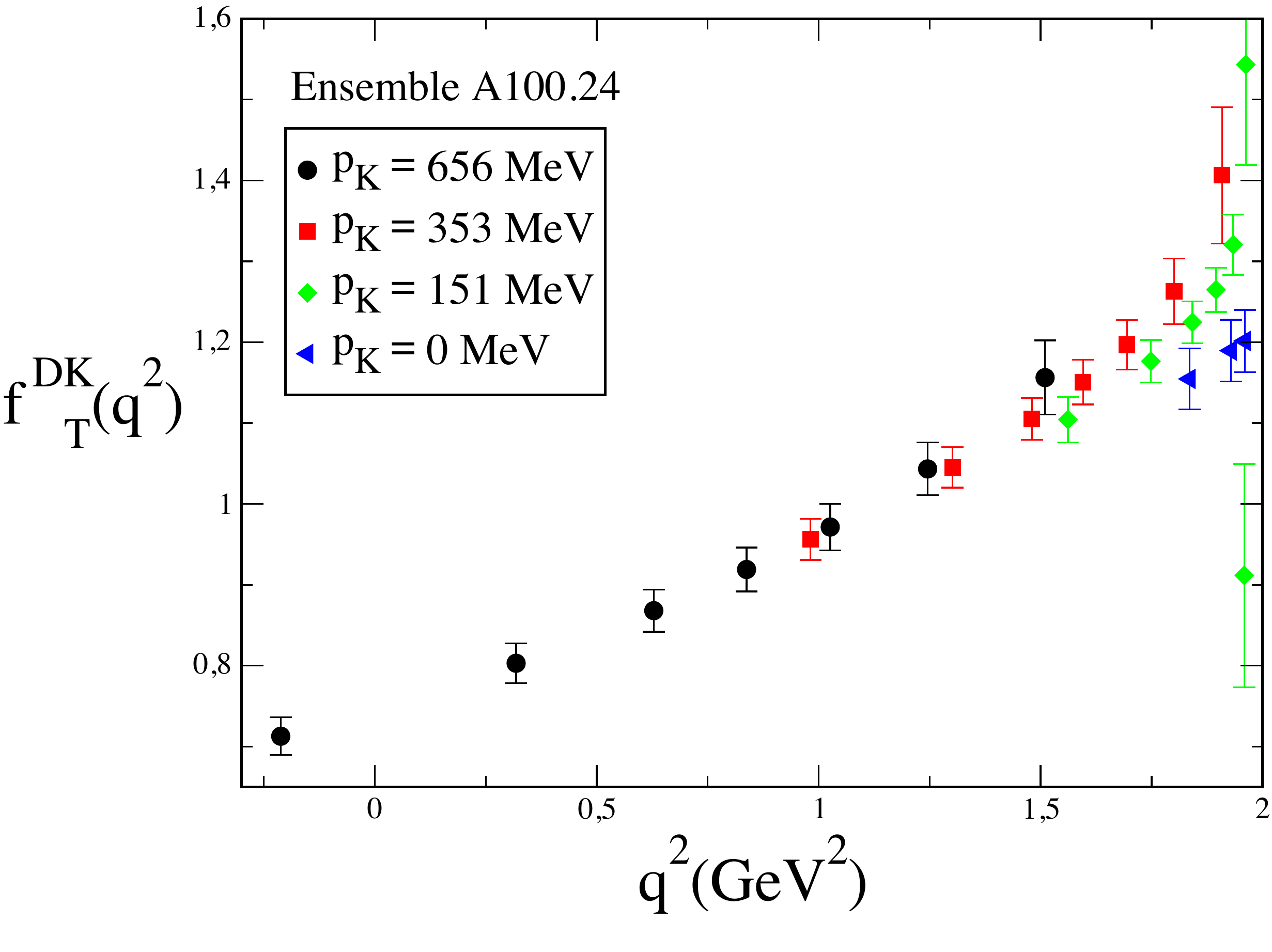}
}
\vspace{-0.75cm}
\caption{\it Momentum dependence of the tensor $D \to \pi$ form factor $f_T^{D\pi}$ (left panel) and of the tensor $D \to K$ form factor $f_T^{D K}$ (right panel) in the case of the gauge ensemble A100.24. Different markers and colors distinguish different values of the child meson momentum. The simulated masses are $M_\pi \simeq 500$ MeV, $M_K \simeq 639$ MeV and $M_D \simeq 2042$ MeV. The lattice spacing and spatial size are $a \simeq 0.0885$ fm and $L \simeq 2.13$ fm, respectively.}
\label{fig:fishbone}
\end{figure}

After a small interpolation of our lattice data to the physical values of the strange and charm quark masses, $m_s^{phys}(2\,\rm{GeV})=99.6\,(4.3)\,$MeV and $m_c^{phys}(2\,\rm{GeV})=1.176\,(39)\,$GeV taken from Ref.~\cite{Carrasco:2014cwa}, we determine the tensor form factor $f_T^{D \to \pi(K)}(q^2)$ for each gauge ensemble and for each choice of parent and child meson momenta. 
The momentum dependencies of the tensor form factors are illustrated in Fig.~\ref{fig:fishbone}, where different markers and colors correspond to different values of the child meson momentum for the ensemble A100.24.
Therefore, if the Lorentz-covariant decomposition (\ref{eq:T_final}) were adequate to describe the lattice data, the extracted form factors would depend only on the squared 4-momentum transfer $q^2$ (and on the parent and child meson masses).
This is not the case and an extra dependence on the value of the child (or parent) meson momentum is clearly visible in Fig.~\ref{fig:fishbone}.

The decomposition (\ref{eq:T_final}) fails to describe our data because, consistently with the Lorentz symmetry, the tensor form factor is assumed to depend only on Lorentz-invariants.
As is well known, however, the lattice breaks Lorentz symmetry and it is invariant only under discrete rotations by multiple of $90^\circ$ in each direction of the Euclidean space-time. Therefore, the form factors may depend also on hypercubic invariants.
In Ref.~\cite{Lubicz:2017syv}, where we observed for the first time the breaking of the Lorentz symmetry in the vector and scalar semileptonic form factors of the $D \to \pi(K)$ transitions, we have proposed a method for the subtraction of these effects, which will be applied to the case of the tensor form factor in the next Section. 

\begin{SCfigure}[1.0][htb!]
\centering
\includegraphics[width=8cm]{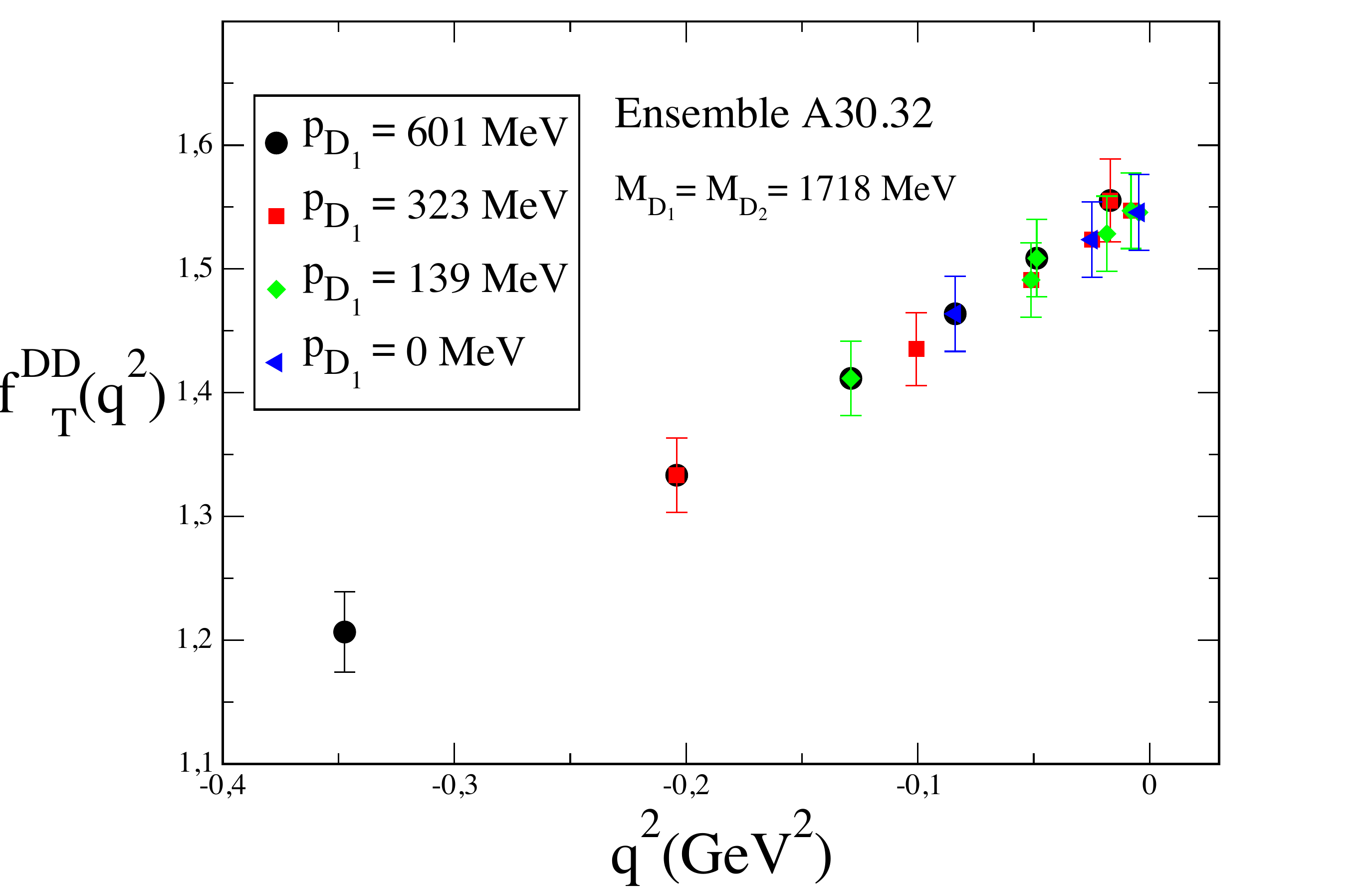}
\caption{\it Momentum dependence of the elastic tensor form factor in which the parent and child mesons are two charmed PS mesons, $D_1$ and $D_2$, with degenerate masses equal to $1718$ MeV in the case of the gauge ensemble A30.32. Different markers and colors distinguish different values of the child meson momentum.\newline \newline \newline \newline \protect\phantom{$\cdots$}}
\label{fig:DtoD}
\end{SCfigure}

Before closing this Section, we remind that in Ref~\cite{Lubicz:2017syv} it was argued that the hypercubic artefacts may be governed by the difference between the parent and the child meson masses. Such an indication is confirmed in the present case by the results shown in Fig.~\ref{fig:DtoD}, where the elastic tensor form factor between two PS-mesons with a mass close to a physical D-meson has been considered. The momentum dependencies of the corresponding form factors show no evidence of hypercubic effects within the statistical uncertainties.

\section{Global fit}\label{sec-2}

We now closely follow the strategy presented in Ref.~\cite{Lubicz:2017syv} for the vector and scalar form factors specializing it to the tensor case. Although the specific Lorentz structure will have important consequences in the decomposition of the matrix elements, the general arguments are the same, and thus we refer the interested reader to the more detailed discussion presented in Sec.~5 of Ref~\cite{Lubicz:2017syv}.
A possible way to describe the observed hypercubic effects is to address them directly on the tensor matrix elements.
Thus, we start by considering the following decomposition:
 \be
    \left\langle {P\left( {{p_P}} \right)} \right|{\widehat{T}_i^E}\left| {D\left( {{p_D}} \right)} \right\rangle  = {\left\langle {{\widehat{T}_i^E}} \right\rangle _{{\text{Lor}}}} + {\left\langle {{\widehat{T}_i^E}} \right\rangle _{{\text{Hyp}}}} ~ ,
    \label{eq:tensor_decomposition}
 \ee
where the suffix $E$ emphasize that this relations are written in the Euclidean space. 
In Eq.~(\ref{eq:tensor_decomposition}) $\braket{\widehat{T}_i^E}_{\rm Lor}$ is the Lorentz-covariant term
 \be
    \braket{\widehat{T}_i^E}_{\rm Lor} = \frac{2}{{{M_D} + {M_P}}}\left[ {{p^4}_P{p^i}_D - {p^i}_P{p^4}_D} \right]{f_T}\left( {{q^2}} \right) ~ ,
    \label{eq:tensor_Lorentz}
 \ee
while $\braket{\widehat{T}_i^E}_{\rm hyp}$ is given by
 \be
    \braket{\widehat{T}_i^E}_{\rm hyp} = {a^2}\frac{2}{{{M_D} + {M_P}}}\left\{ {\left[ {{{\left( {{p^4}_P} \right)}^3}{p^i}_D - {{\left( {{p^i}_P} \right)}^3}{p^4}_D} \right]{H_1} + \left[ {{p^4}_P{{\left( {{p^i}_D} \right)}^3} - {p^i}_P{{\left( {{p^4}_D} \right)}^3}} \right]{H_2}} \right\}
    \label{eq:tensor_hypercubic}
 \ee
and the quantities $H_i$ ($i = 1, 2$) are additional hypercubic form factors. In Eqs.~(\ref{eq:tensor_Lorentz}-\ref{eq:tensor_hypercubic}) the suffix $4$ indicate the time component of a Eucledian vector which is related to the time component of the same vector in Minkowski space via $p^4=i p^0$.

Eq.~(\ref{eq:tensor_hypercubic}) is the most general structure, up to order $\mathcal{O}(a^2)$, that transforms properly under hypercubic rotations, is antisymmetric under exchange of two space-time indices, and is built with fourth powers of the components of the parent and child momenta $p^\mu_D$ and $p^\mu_P$. 
The Lorentz-invariance breaking effects are encoded in the two structures proportional to the hypercubic form factors $H_i$, which, we assume, depend only on $q^2$ (and on the parent and child meson masses). 
For the $H_i$ form factors we adopt a simple polynomial form in terms of the $z$ variable \cite{Boyd:1995cf,Arnesen:2005ez}
 \be
    H_i(z) = d_0^i + d_1^i z + d_2^i z^2 ~ ,
    \label{eq:Hi}
 \ee
where the coefficients $d_{0,1,2}^i$ are treated as free parameters. 
\noindent
The hypercubic structure~(\ref{eq:tensor_hypercubic}) cannot be determined and subtracted separately at the level of each gauge ensemble, but it can be fitted by studying simultaneously all the data for the 15 ETMC gauge ensembles. Thus we performed a global fit considering the dependencies on $q^2$, $m_\ell$ and $a^2$ of the Lorentz form factor $f_{T}$ as well as the $q^2$ and $m_\ell$ dependencies of the hypercubic form factors.

\begin{figure}[htb!] 
\centering
\makebox[\textwidth][c]{
\includegraphics[width=6.75cm,clip]{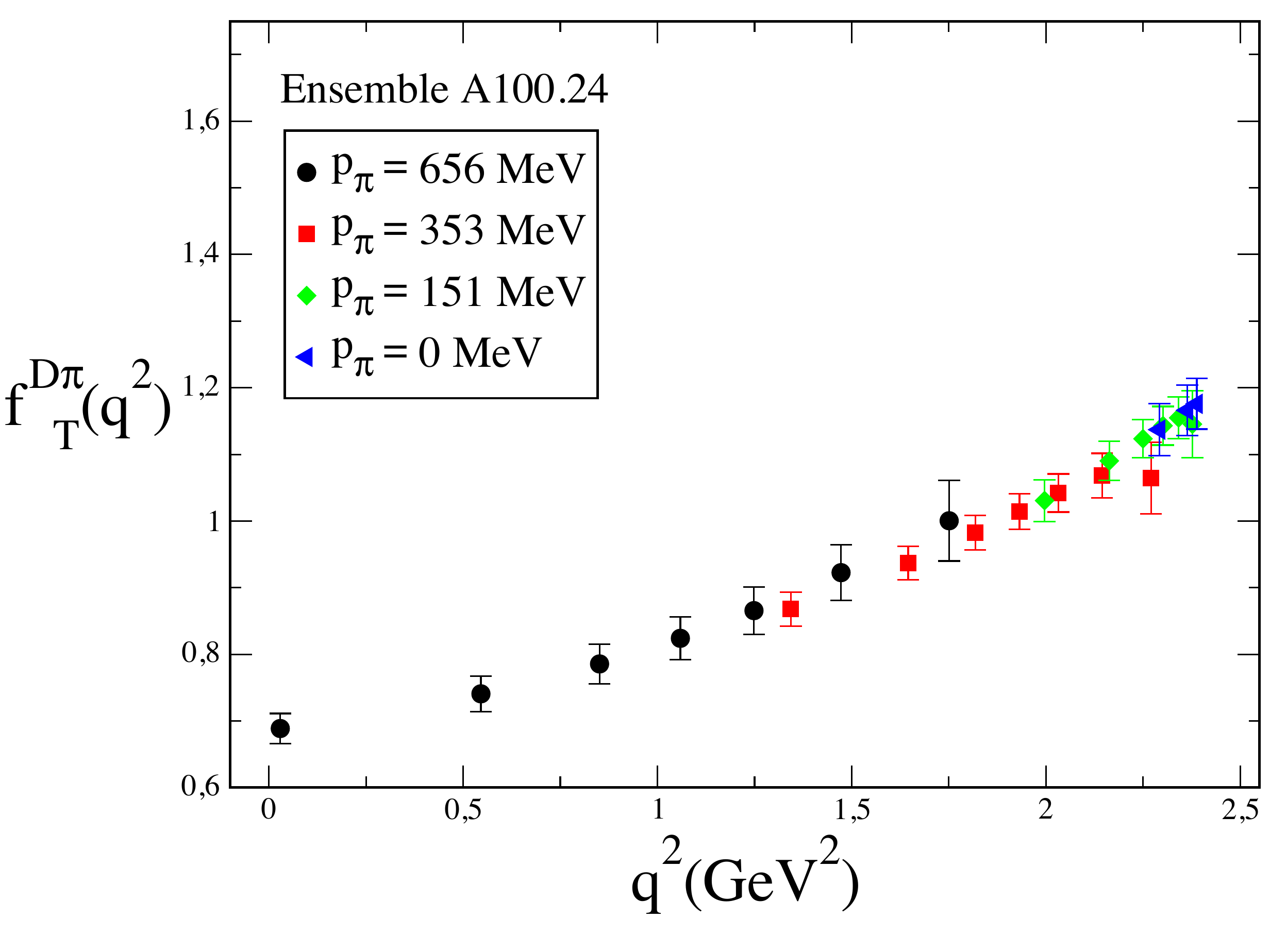}
\includegraphics[width=6.75cm,clip]{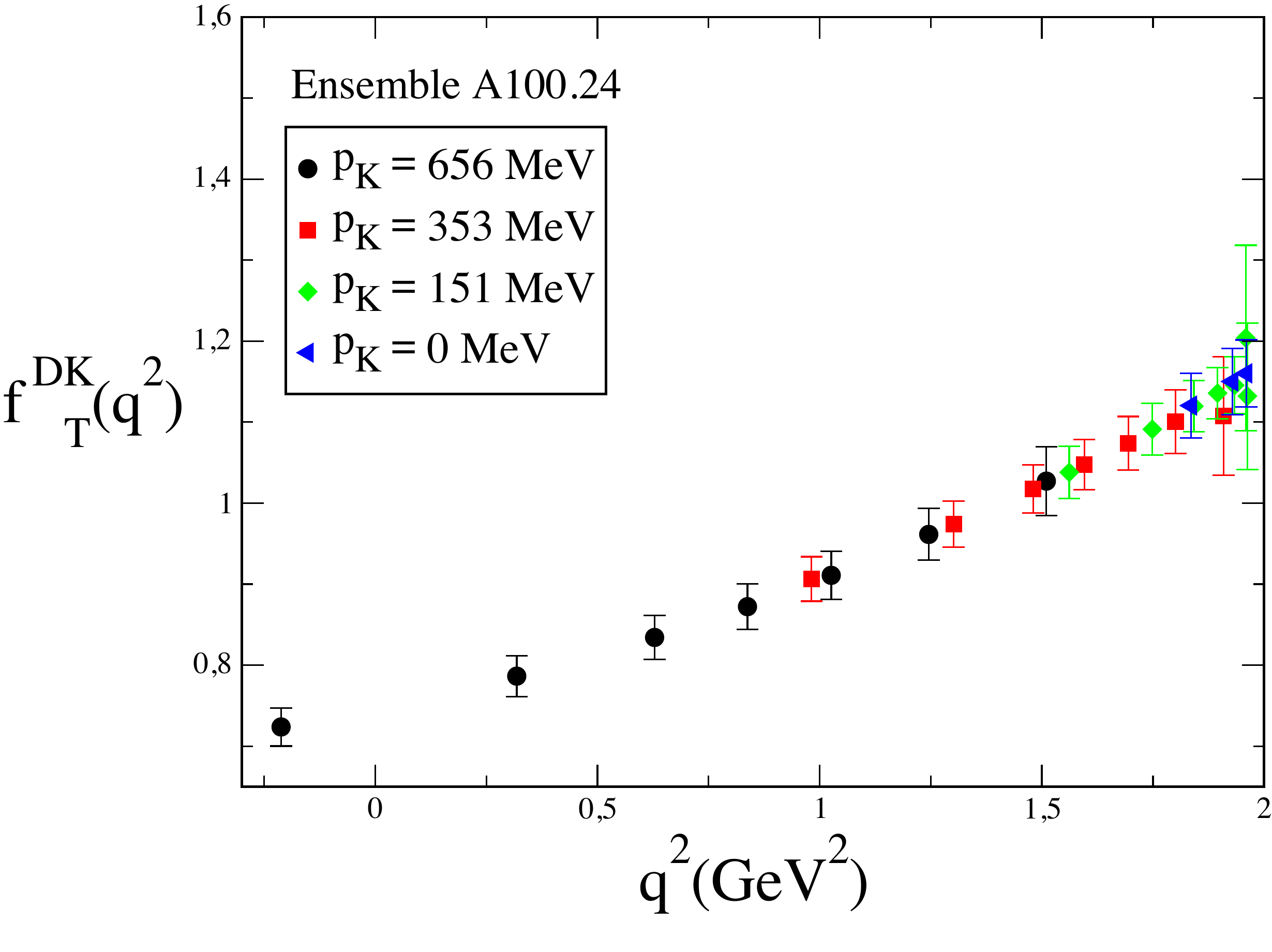}
}
\vspace{-0.75cm}
\caption{\it Same ensembles showed in Fig.~\ref{fig:fishbone} after removing the hypercubic effects determined by the global fit.}
\label{fig:corrected}
\end{figure}

For the form factor $f_{T}(q^2, a^2)$ we have adopted the modified z-expansion of Ref.~\cite{Bourrely:2008za}, viz.
 \be
   \label{eq:z-exp_fT}
    f_T^{D \to \pi(K)}(q^2, a^2) = \left[ f^{D \to \pi(K)}(0, a^2) + c^{\pi(K)}(a^2) (z - z_0)
                                                   \left(1 + \frac{z + z_0}{2} \right) \right] {\Large \mbox{/}} 
                                                   \left[ 1 - q^2 / \left( M^{\pi(K)}_T \right)^2 \right] ~ ,
 \ee
where we assume for the coefficients $c^{\pi(K)}(a^2)$ a linear dependence on $a^2$, while the pole masses $M^{\pi(K)}_T$ are treated as free parameters in the fitting procedure.\\
\noindent
For the vector form factor at zero 4-momentum transfer, $f^{D \to \pi(K)}(0, a^2)$, we use the following Ansatz
\be
    \label{eq:ChLim}
    f^{D \to \pi(K)}(0, a^2) =  \left( M_D + M_{\pi (K)} \right) F^{\pi(K)} \left[ 1 + A^{\pi(K)} \xi _\ell \log {\xi _\ell} + 
                                           b_1 \xi _\ell + D a^2 \right] ~,
\ee
where the coefficients $F^{\pi(K)}$, $b_1$ and $D$ are treated as free parameters in the fitting procedure. The chiral-log coefficient $A^{\pi(K)}$ is either left as a free parameter or put equal to zero. The difference between these two Ansatze is used to estimate the systematic uncertainty relative to the chiral extrapolation.
In Fig.~\ref{fig:corrected} we show the same form factors given in Fig.~\ref{fig:fishbone} after the hypercubic contributions determined by the global fit have been subtracted from $\braket{\widehat{T}_i}$. It can be seen that the tensor form factors depend now only on the $4-$momentum transfer $q^2$. 

\begin{figure}[htb!] 
\centering
\makebox[\textwidth][c]{
\includegraphics[width=6.75cm,clip]{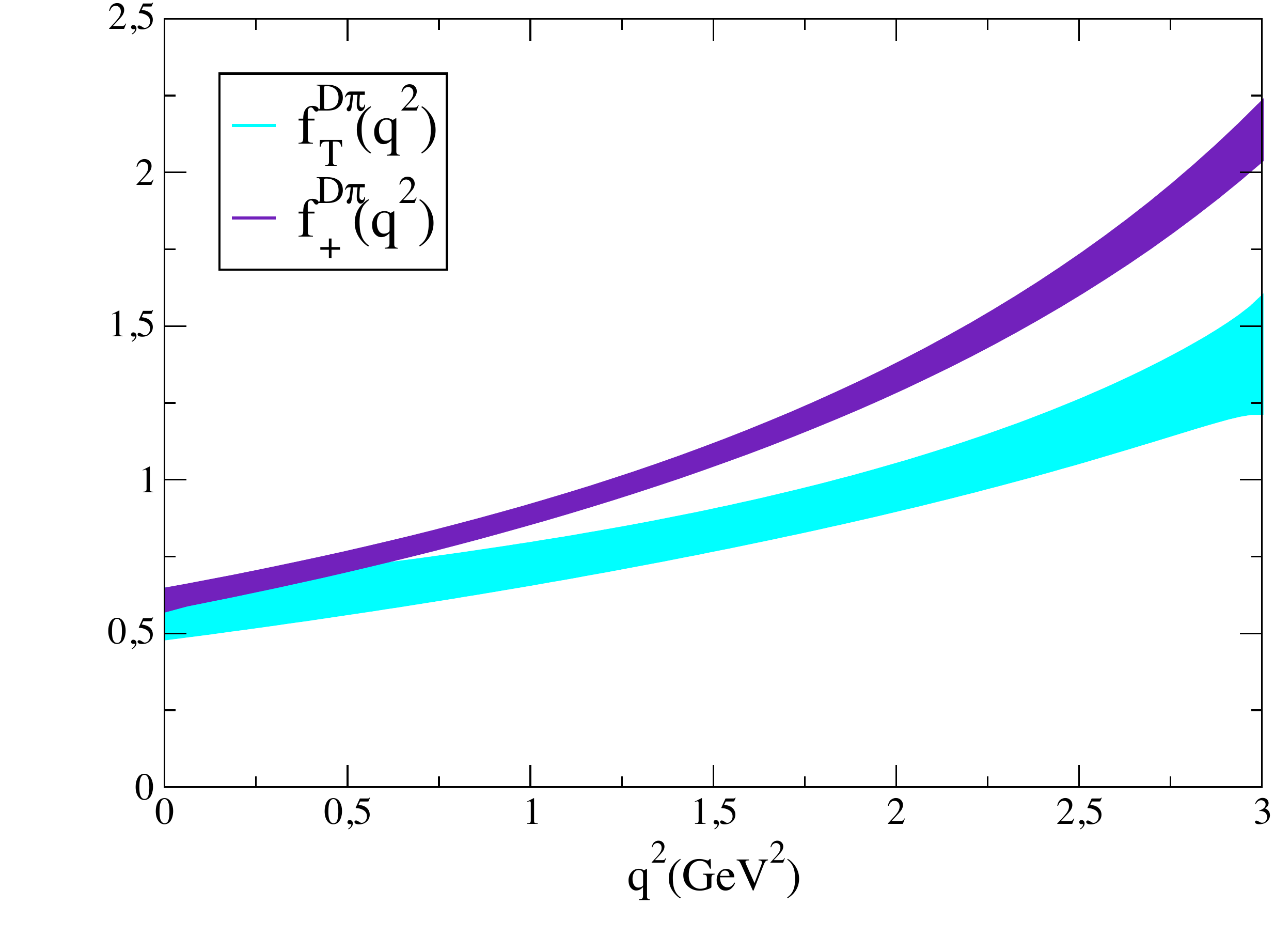}
\includegraphics[width=6.75cm,clip]{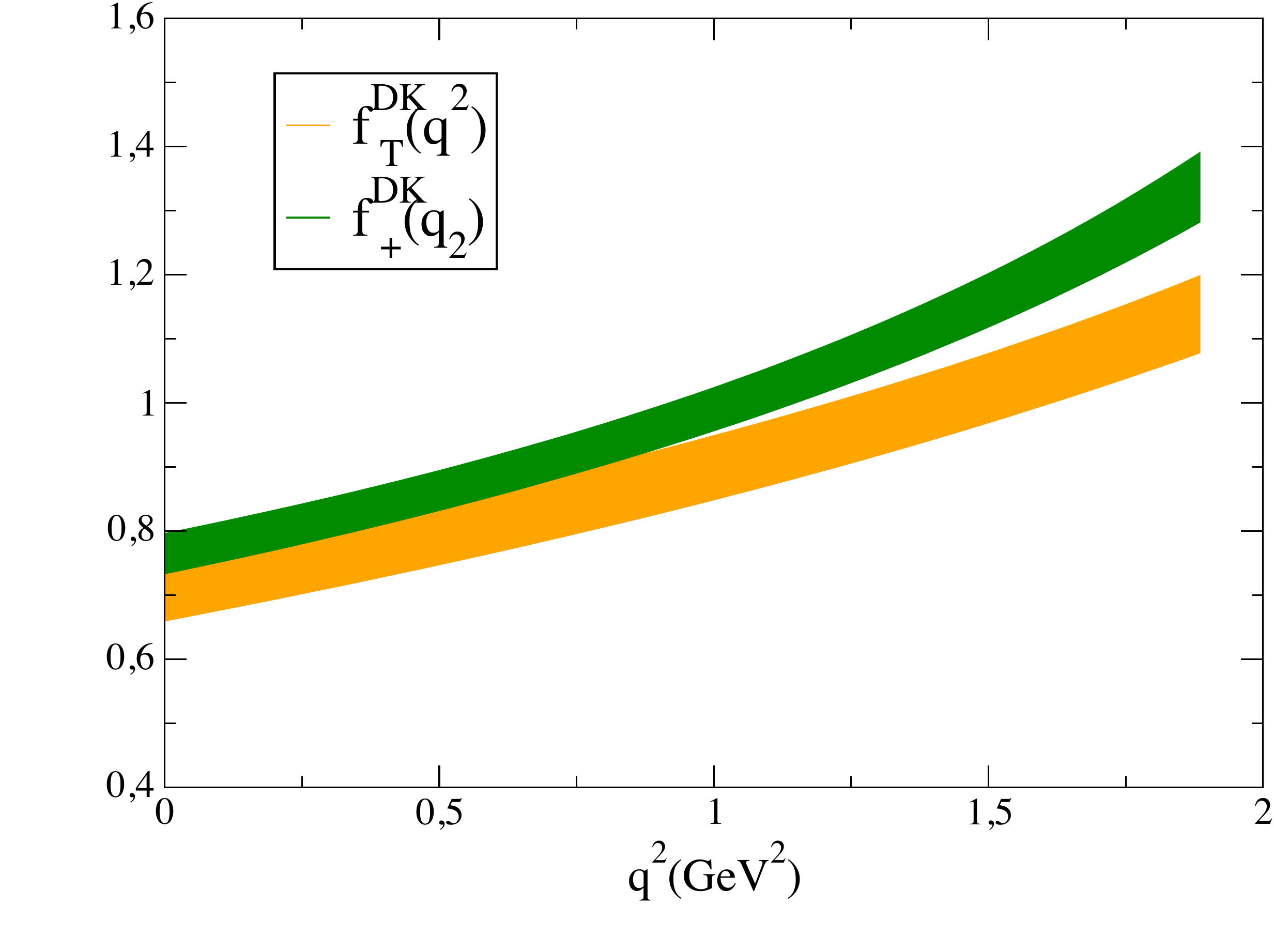}
}
\vspace{-0.75cm}
\caption{\it Momentum dependencies of the Lorentz-invariant form factor $f_T(q^2)$, calculated in this work, and $f_+(q^2)$, from Ref.~\cite{Lubicz:2017syv}, for the $D \to \pi$ (left panel) and $D \to K$ (right panel) transitions. Both form factors are extrapolated to the physical pion mass and to the continuum and infinite volume limits. The bands correspond to the total (statistical + systematic) uncertainty at the level of one standard deviation.}
\label{fig:physicalFormFactors}
\end{figure}

The momentum dependencies of the physical Lorentz-invariant tensor form factors, extrapolated to the physical pion mass and to the continuum and infinite volume limits, are shown in Fig.~\ref{fig:physicalFormFactors} as a cyan(orange) band for the $D \to \pi(K)$ transition. 
In Fig.~\ref{fig:physicalFormFactors} the tensor form factors are compared with the corresponding vector ones $f^{D \to \pi(K)}_{+}(q^2)$ extracted from the same gauge ensembles in Ref.~\cite{Lubicz:2017syv}.
Finally, at zero 4-momentum transfer our preliminary results in the $\overline{\rm MS}(2~\mbox{GeV})$ scheme are
 \bea
        \qquad \qquad \qquad \qquad \quad
    f_T^{D \to \pi}(0) = 0.563 ~ (67)_{\rm stat} ~ (42)_{\rm syst} = 0.563 ~ (80) ~ ,\\[2mm]
    f_T^{D \to K}(0) = 0.706 ~ (45)_{\rm stat} ~ (14)_{\rm syst} = 0.706 ~ (47) ~ ,
    \label{form_factor_at_zero}
 \eea
where the systematic error includes the uncertainties due to the chiral extrapolation, the discretization effects and the RC ${\cal{Z}}_T$.

\section*{Acknowledgements}
We thank our colleagues of the ETMC for fruitful discussions.
We gratefully acknowledge the CPU time provided by PRACE under the project PRA067 and by CINECA under the initiative INFN-LQCD123 on the BG/Q system Fermi at CINECA (Italy).

\bibliography{lattice2017}

\end{document}

%% file: definitions.tex
\newcommand{\be}{\begin{equation}}
\newcommand{\ee}{\end{equation}}
\newcommand{\bea}{\begin{eqnarray}}
\newcommand{\eea}{\end{eqnarray}}
\newcommand{\bi}{\begin{itemize}}
\newcommand{\ei}{\end{itemize}}

\newcommand{\bspl}{\begin{split}}
\newcommand{\espl}{\end{split}}

\newcommand{\MeV}{\,\mathrm{MeV}}

\newcommand{\fm}{\,\mathrm{fm}}

\def\mev{{\rm MeV}}
\def\gev{{\rm GeV}}
\def\tev{{\rm TeV}}
\def\fm{{\rm fm}}
 
\def\fm{\mathrm{fm}}
\def\ev{\mathrm{e\kern-0.1em V}}
\def\kev{\mathrm{ke\kern-0.1em V}}
\def\mev{\mathrm{Me\kern-0.1em V}}
\def\gev{\mathrm{Ge\kern-0.1em V}}
\def\tev{\mathrm{Te\kern-0.1em V}}